\documentclass[twoside]{ilcws10}
\usepackage[latin1]{inputenc}
\usepackage[dvips]{graphicx,epsfig,color}
\usepackage{wrapfig,rotating}
\usepackage{amssymb,amsmath,array}

\usepackage{cite}
\usepackage{url}

\usepackage[tight]{subfigure}

\DeclareMathOperator{\erf}{erf}                 
\DeclareMathOperator{\sign}{sign}               
\DeclareMathOperator{\GeV}{GeV}

\newcommand{\artitle}[1]{}
\newcommand{\artitlekeep}[1]{``#1''}
\newcommand{\sub}[1]{\ensuremath{_{\mathrm{#1}}}} 
\newcommand{\super}[1]{\ensuremath{^{\mathrm{#1}}}} 
\newcommand{\mWgen}{\ensuremath{m\sub{W}\super{gen}=80.419\,\GeV}}

\newcommand{\pz}{\ensuremath{p\sub{z}}}               
\newcommand{\pzgamma}{\ensuremath{p\sub{z,\gamma}}}         
\newcommand{\pzgammagen}{\ensuremath{p\sub{z,\gamma}\super{gen}}} 
            
\newcommand{\EISR}{\ensuremath{E\sub{ISR}}}           
\newcommand{\Emax}{\ensuremath{E\sub{max}}}

\begin{document}
\title{
Consideration of Photon Radiation in Kinematic Fits for Future $e^+e^-$ Colliders} 
\author{Moritz Beckmann$^{1,2}$, Benno List$^2$ and Jenny List$^1$
\vspace{.3cm}\\
1 - DESY, Notkestr. 85, 22607 Hamburg, Germany
\vspace{.1cm}\\
2 - University of Hamburg, Institute for Experimental Physics\\
    Luruper Chaussee 149, 22761 Hamburg, Germany
}

\maketitle

\begin{abstract}
Kinematic fitting is an important tool to improve the resolution in high-energy 
physics experiments. 
At future $e^+ e^-$ colliders, photon radiation parallel to the beam
carrying away large amounts of energy and momentum will become a 
challenge for kinematic fitting. 
A photon with longitudinal momentum 
$\pzgamma\,(\eta)$ is introduced, which is parametrized such that
$\eta$ follows a normal distribution.
In the fit, $\eta$ is treated as having a measured value of zero,
which corresponds to $\pzgamma=0$.
As a result,
fits with constraints on energy and momentum conservation
converge well even in the presence of a highly energetic photon,
while the resolution of fits without such a photon
is retained.
A fully simulated and reconstructed 
$e^+ e^- \rightarrow q \bar q q \bar q$ 
event sample at $\sqrt{s}=500\,\GeV$
is used to investigate the performance of this method
under realistic conditions,
as expected at the International Linear Collider.
\end{abstract}


\section{Introduction}
\label{sec:intro}

Radiation of photons at angles so small that they escape along 
the beam pipe is usually not taken into account in kinematic fits. 
At previous $e^+ e^-$ colliders such as LEP, the losses due to photon 
radiation were acceptable \cite{lep}. At future facilities 
such as the International Linear Collider (ILC) or 
the Compact Linear Collider (CLIC), photon 
radiation will be much stronger due to higher center-of-mass energies and 
stronger focussing of the beams.

This paper presents a novel method to take the energy and longitudinal
momentum of photon radiation into account in kinematic fits. 
A priori information 
about the momentum spectrum of photon radiation is used to 
treat the photon's momentum as a measured parameter in the fit. 
As a test case, the production of $W^{+}W^{-}$/$Z^0Z^0$ pairs
decaying to light quark jets at the ILC is considered,
with fully simulated Monte Carlo events as reconstructed by the
International Large Detector (ILD) \cite{bib:ild}. A more detailed
description of the underlying concept can be found in \cite{bib:isr-paper}
and a more detailed description of the method and its application tests in
\cite{bib:diplomarbeit}.


\section{Representation of the photon}
\label{sec:representation}

The simplest method to cope with highly energetic photons escaping the detector
along the beam pipe in a constrained kinematic fit is therefore to drop the energy
and longitudinal momentum conservation constraints, thus losing two degrees of freedom.

Here the photon is treated as a particle with a measured momentum of zero and
an uncertainty derived from its known momentum spectrum.
For this purpose, the photon momentum \pzgamma\ is transformed into a quantity $\eta$
which follows a Gaussian distribution \cite{bib:isr-paper}:

\begin{equation}
  \pzgamma\,(\eta) = \sign(\eta)\,\Emax \, \left [ \erf(|\eta|/\sqrt{2}) \right ]^{\frac{1}{\beta}}
  \label{eq:parametrization}
\end{equation}

$\Emax$ is the maximum possible energy for a single ISR photon. The exponent $\beta$ is given by

\begin{equation}
  \label{eq:betavalue}
   \beta = \frac{2 \alpha}{\pi}\, \left ( \ln \frac{s}{m\sub e^2} - 1\right ),
\end{equation}
which corresponds to $\beta = 0.1235$ for $\sqrt s = 500\,\GeV$.

Then the photon will be treated as if it had a measured value of
$\eta\sub{meas}=0$. By this procedure, the a priori knowledge of the
photon's energy spectrum (in particular the fact that it is
negligibly small in most cases) is used, and all energy and momentum
constraints can be applied.

\section{Performance tests}
\label{sec:performance}

The method described above is applied to the process 
$e^+ e^- \to W^+ W^- \to 4\,{\mathrm{jets}}$ events.
The fraction of successful fits, the width and the shift
of the reconstructed $W^\pm$ mass peak are used 
to compare the performance of the various 
kinematic fit variants.

\subsection{Data set}
\label{subsec:dataset}

The analysis sample $e^+ e^- \rightarrow u\bar{d}d\bar{u}$
was generated using 
the matrix element generator WHIZARD \cite{bib:whizard},
which takes into account all Feynman diagrams
leading to a given final state, including interference terms.

The initial state radiation is also simulated by WHIZARD, the beamstrahlung spectrum is 
simulated with GUINEA-PIG \cite{bib:guineapig}. For this calculation the nominal
beam parameter set of the ILC was assumed. 

A full simulation of the ILD detector 
\cite{bib:ild} is performed by the GEANT based simulation program 
MOKKA \cite{bib:mokka}.
In the event reconstruction,
which is implemented as part of the software package MarlinReco 
\cite{bib:marlinreco}, the tracks are matched to the calorimeter 
clusters by the Pandora particle flow algorithm \cite{bib:pandora} and the 
resulting reconstructed particles are forced into four jets by the 
Durham algorithm \cite{bib:durham}. 

In order to investigate the influence of ISR and beamstrahlung
on the performance of the kinematic fit, the results of all three fits
are given for the complete event sample as well as for three subsamples
with different amounts of ISR energy: $\EISR<5\,\GeV$; $5\,\GeV \le \EISR \le 30\,\GeV$ and
$\EISR> 30\,\GeV$.

\subsection{Evaluation method}
\label{subsec:fits}

In order to investigate the performance of the proposed method, kinematic fits 
are applied to the four jets in the events 
from  the test sample, comparing the event hypotheses ``4~jets'' ($4j$) 
and ``4~jets + 1~photon'' ($4j+\gamma$). 
Both event hypotheses are fitted with  
five constraints ($5C$-fit): conservation of energy, conservation of the three 
momentum components and equal di-jet masses. 
In addition, the events are fitted also using only the three constraints
($3C$-fit) that are not affected by the presence of photon radiation,
i.e. conservation of the transverse momentum components and the equal
mass constraint.

Both values $\pzgamma=0$ and the missing \pz\ from the event
$\pzgamma = p\sub{z, miss}$
are considered as starting values for the photon momentum
in the kinematic fit, and the result with the better fit performance is chosen.

\begin{figure}[hbt]
   \epsfig{file=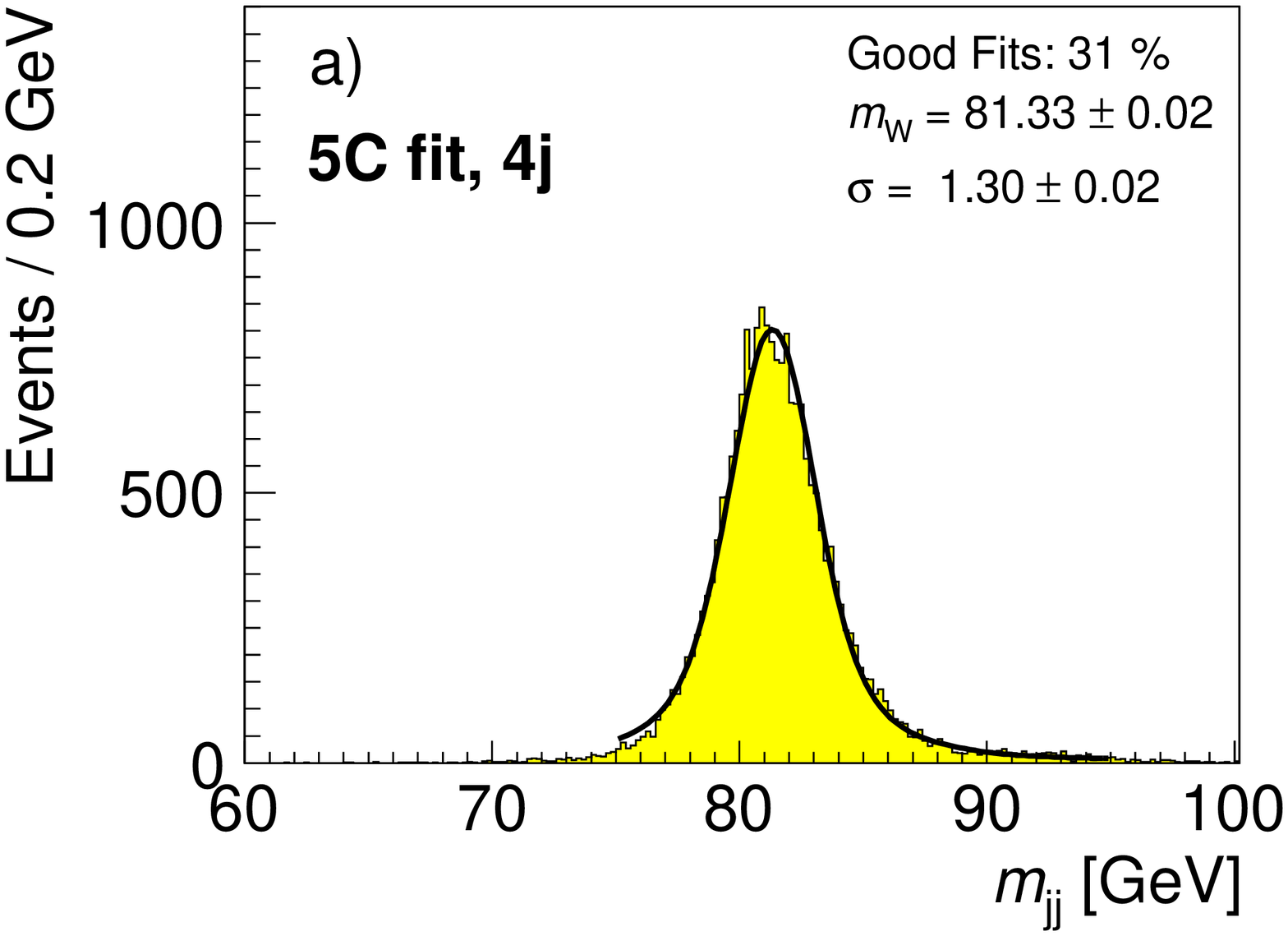,width=0.475\textwidth}
   \epsfig{file=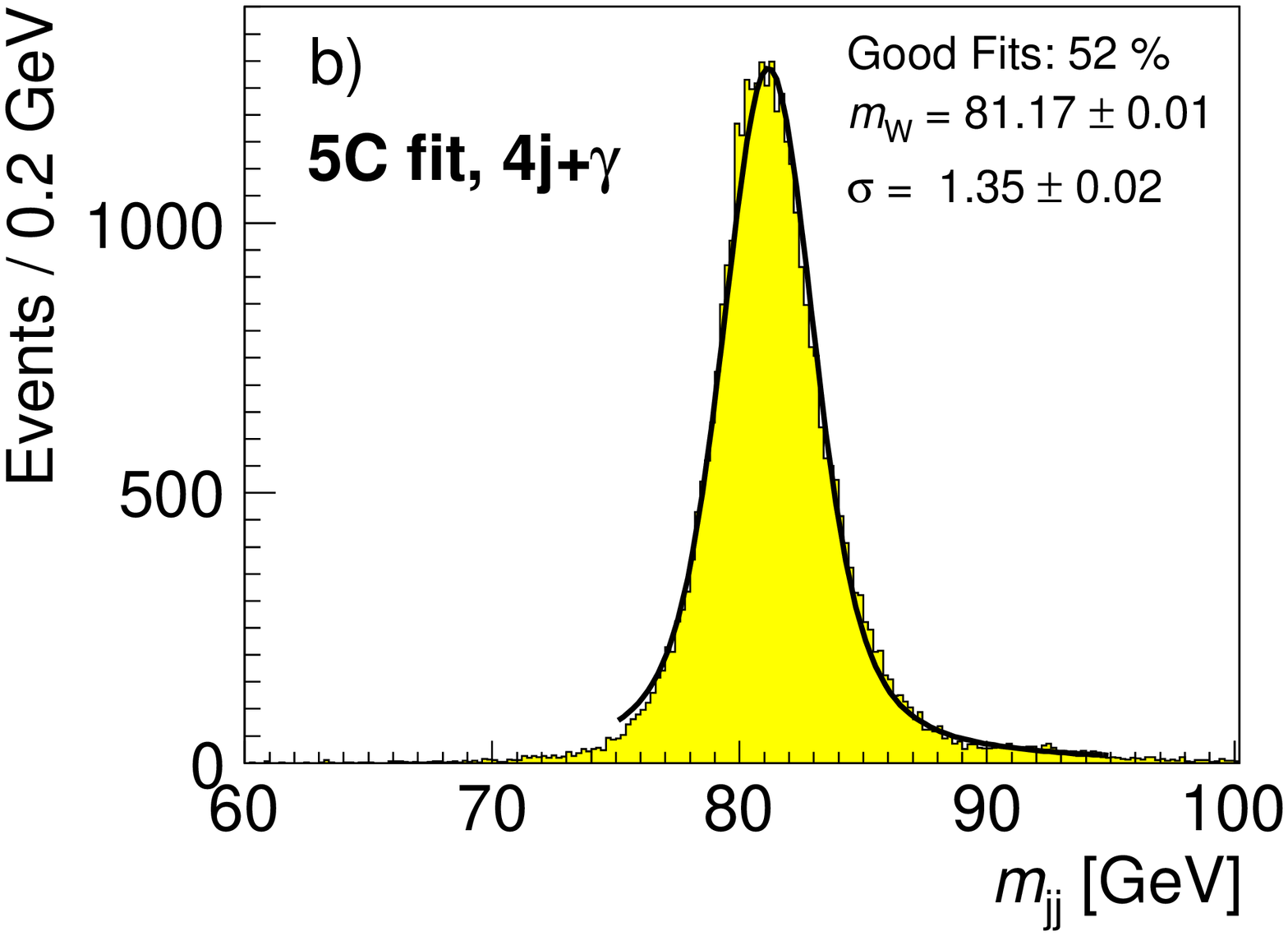,width=0.475\textwidth}
   \caption{Invariant di-jet masses $m\sub{jj}$ for the Monte Carlo 
        sample described in the text:
        a) the di-jet masses $m\sub{jj}$ for the $5C$ fit under a $4j$ hypothesis;
        b) $m\sub{jj}$ for the $5C$ fit under a $4j+\gamma$ hypothesis.
   \label{fig:dijetmasses}
        }
\end{figure}

The fits are compared in terms of the following three quantities: the
fraction of fits with a fit probability $> 0.001$, the difference
$\Delta m\sub{W} = m\sub{W} - m\sub{W}\super{gen}$ between
the peak position of the reconstructed $W^\pm$ mass spectrum and the input mass \mWgen,
and the Gaussian width of the peak. The latter two parameters have been
determined from a fit to the mass spectrum which takes into account the
natural width of the $W^\pm$ and the small fraction of $ZZ$ events in the sample
\cite{bib:isr-paper}. Fig.~\ref{fig:dijetmasses} shows the the invariant di-jet 
masses before and after the kinematic fit for the complete sample, including ISR and beamstrahlung.

If large amounts of energy are missing, the fitted jet energies have 
to be larger than the measured ones to fulfill energy conservation. 
Consequently,  di-jet masses are shifted to higher values
and thus a larger $\Delta m\sub{W}$ is obtained. 
Due to imperfections of the lineshape fit,
a nonzero value of $\Delta m\sub{W}$ is to be expected,
for which a correction would be applied in a real analysis.
However, if this mass shift depends on the amount of 
energy from ISR and beamstrahlung,
it leads to a broadening of the signal and thus a loss of
resolution; in addition, systematic uncertainties arise
from the description of the ISR and in particular the beamstrahlung
energy spectrum.
Therefore, a mass shift that is independent of the amount
of energy lost to ISR and beamstrahlung is desirable.

\subsection{Results}
\label{subsec:results}

Tab.~\ref{tab:results} summarizes the results of our tests.
It lists the fraction of good fits, the mass shift and
the width of the Gaussian part of the peak
for the complete sample, as well as the three subsamples with
different amounts of missing energy due to ISR photons.
The results are given for the average of the di-jet masses 
before a kinematic fit, using the $3C$ jet pairing,
as well as the di-jet mass after applying a $3C$ fit or
a $5C$ fit without or with an ISR photon.
The results are reported for the case where the effect from 
beamstrahlung has been excluded,
and for the realistic case where effects from ISR and beamstrahlung are 
fully taken into account.

\begin{table}[tb]
\begin{tabular}{|l|l|ccc|ccc|}
   \hline
   Subsample  & Constraints, & \multicolumn{3}{|c|}{ISR only} & 
                                    \multicolumn{3}{|c|}{Full Photon Spectrum} \\
   (Fraction) & Hypothesis   & Good        & $\Delta m\sub{W}$ & $\sigma\sub{W}$ & 
                                    Good        & $\Delta m\sub{W}$ & $\sigma\sub{W}$  \\
              &              & fits $[\%]$ & $[\GeV]$          & $[\GeV]$        & 
                                    fits $[\%]$ & $[\GeV]$          & $[\GeV]$  \\
   \hline
   \hline
   All events                   & ---             & $55\,\%$ & $+0.78$ & $2.05$ & 
                                                         $55\,\%$ & $+0.78$ & $2.05$ \\
   $(100\,\%)$                  & $3C, 4j$        & $55\,\%$ & $+0.82$ & $2.06$ & 
                                                         $55\,\%$ & $+0.82$ & $2.06$ \\
                                & $5C, 4j$        & $42\,\%$ & $+0.67$ & $1.21$ & 
                                                         $31\,\%$ & $+0.91$ & $1.30$ \\
                                & $5C, 4j+\gamma$ & $54\,\%$ & $+0.53$ & $1.25$ & 
                                                         $52\,\%$ & $+0.75$ & $1.35$ \\
   \hline
   $\EISR<5\,\GeV$             & ---             & $56\,\%$ & $+0.80$ & $2.04$ & 
                                                         $56\,\%$ & $+0.80$ & $2.04$ \\
   $(75\,\%)$                   & $3C, 4j$        & $56\,\%$ & $+0.85$ & $2.06$ & 
                                                         $56\,\%$ & $+0.85$ & $2.06$ \\
                                & $5C, 4j$        & $53\,\%$ & $+0.63$ & $1.19$ & 
                                                         $40\,\%$ & $+0.86$ & $1.27$ \\
                                & $5C, 4j+\gamma$ & $55\,\%$ & $+0.49$ & $1.24$ & 
                                                         $54\,\%$ & $+0.69$ & $1.31$ \\
   \hline
   $5 \GeV \le \EISR$           & ---             & $54\,\%$ & $+0.79$ & $2.07$ & 
                                                         $54\,\%$ & $+0.79$ & $2.07$ \\
   $\le 30\,\GeV$               & $3C, 4j$        & $54\,\%$ & $+0.84$ & $2.08$ & 
                                                         $54\,\%$ & $+0.84$ & $2.08$ \\
   $(11\,\%)$                             & $5C, 4j$        & $15\,\%$ & $+1.68$ & $1.25$ & 
                                                         $12\,\%$ & $+2.19$ & $1.29$ \\
                                & $5C, 4j+\gamma$ & $53\,\%$ & $+0.71$ & $1.27$ & 
                                                         $50\,\%$ & $+1.07$ & $1.51$ \\
   \hline  
   $\EISR > 30\,\GeV$           & ---             & $53\,\%$ & $+0.59$ & $1.99$ & 
                                                         $53\,\%$ & $+0.59$ & $1.99$ \\
   $(13\,\%)$                   & $3C, 4j$        & $53\,\%$ & $+0.66$ & $1.99$ & 
                                                         $53\,\%$ & $+0.66$ & $1.99$ \\
                                & $5C, 4j$        & $0\,\%$  & ---     &    --- & 
                                                         $0\,\%$  & ---     &    --- \\
                                & $5C, 4j+\gamma$ & $47\,\%$ & $+0.64$ & $1.21$ & 
                                                         $42\,\%$ & $+0.91$ & $1.38$ \\
   \hline
\end{tabular}
\caption{
  \label{tab:results}
  Results of kinematic fits under various conditions.
  ``ISR only'' refers to the case where the effect of beamstrahlung and beam energy spread is 
  removed from the fit, while ``Full Photon Spectrum'' includes these effects.
  For each fit variation, the fraction of good fits with fit probability $p>0.001$,
  the difference $\Delta m\sub{W}$ between the fitted and generated W mass of \mWgen, 
  and the width of the Gaussian part of the peak is given.
  The rows refer to the results from averaging the measured di-jet masses without a fit
  for events where the 3C fit converges,
  the 3C fit with only transverse momentum and equal-mass constraint,
  the 5C fit under a four jet hypothesis with longitudinal momentum and energy constraints in addition,
  and the 5C fit with an additional ISR photon fit object.
  The subsamples are distinguished by the total energy $\EISR$
  of ISR photons, excluding beamstrahlung.
}
\end{table}

\subsubsection*{Results with ISR only}

A comparison of the fit results
demonstrates the gain in resolution 
achieved by kinematic fitting: The Gaussian $\sigma$, which corresponds to
the di-jet mass resolution, is $\sigma = 2.1\,\GeV$ for the average of the 
two di-jet masses without a kinematic fit
and improves to $\sigma = 1.3 \,\GeV$ if a kinematic fit with five constraints
is used.
A fit with only three constraints does not improve the 
resolution compared to the simple averaging of the unfitted di-jet masses.

The fit with five constraints and no ISR photon cannot be applied to the
subsample with $\EISR > 30\,\GeV$, because fit probabilities above
the cut of $p = 0.001$ are essentially never achieved
due to the too large amounts of missing energy and momentum.
Therefore this subsample, which contains $13\,\%$ of all events, cannot be used
for an analysis.
The $5C$ fit with an ISR photon, on the other hand, achieves almost the same
performance for the two subsamples with $\EISR > 30\,\GeV$ and $\EISR < 5\,\GeV$
in terms of the fraction of good fits ($47\,\%$ vs. $55\,\%$)
as well as in resolution ($\sigma = 1.21\,\GeV$ vs. $1.24\,\GeV$)
with only a small additional bias in the W mass 
($\Delta m\sub{W} = 0.64\,\GeV$ vs. $0.49\,\GeV$).

The sample with moderate ISR energy  $5 \GeV \le \EISR \le 30\,\GeV$,
which comprises $11\,\%$ of the events,
demonstrates that the $5C$ fit without the inclusion of an ISR photon
tends to develop a mass bias. 
This is because the energy carried away by the photon
is falsely attributed to the final state jets, which increases their energy
and thus the invariant mass: The mass bias increases from
$\Delta m\sub{W} = +0.63\,\GeV$ to $+1.68\,\GeV$. At the same time, only
$15\,\%$ of the events yield a good $5C$ fit under the $4j$ hypothesis.
In contrast, the $4j+\gamma$ hypothesis shows the same performance in terms of 
fraction of good fit, mass shift and resolution as for the sample with
small missing energy.

The fact that for all fit hypotheses only about half of the events
have reasonable fit probabilities $p > 0.001$ can be mostly attributed to the 
equal-mass constraint: The resolution for the difference of
the di-jet masses is approximately $4.1\,\GeV$ (twice the resolution for the 
di-jet mass average for the unfitted jets),
which is of similar size as the broadening
of $4.3\,\GeV$ due to the intrinsic W width. 
This indicates that in a real analysis 
the na\"ive equal-mass constraint has to be modified
to take the natural $W$ width into account.
Other factors that reduce the fraction of successful fits are
events from processes other than $W/Z$ boson pair production
and the fact that the jet error parametrization employed in this 
analysis does not include the effects of parton showering.

\subsubsection*{Results with ISR and beamstrahlung}

The right-hand side of Tab.~\ref{tab:results}
shows the results for the case where the effect of both,
ISR and beamstrahlung, is considered.
Because the three subsamples are defined on the basis of
the ISR energy only, the same amount of beamstrahlung is present
in each of them.
A comparison with
the case where only the effect from ISR is considered,
demonstrates that the photon momentum parametrization
Eq.~(\ref{eq:parametrization}) derived from the ISR momentum
spectrum also works quite well in the presence of beamstrahlung,
at least at the level of beamstrahlung that is expected
for the nominal ILC parameter set.

Since beamstrahlung in the Monte Carlo simulation used for this analysis
is simulated solely through a variation of the energy of the incoming 
leptons, no transverse momentum is carried by the beamstrahlung.
Therefore the results for the $3C$ fit and the 
di-jet masses calculated without a kinematic fit do not change
when beamstrahlung effects are considered.

The performance of the $5C$ fit under the $4j$ hypothesis
is significantly reduced when beamstrahlung effects are considered
due to the larger amount of missing energy.
Overall, the fraction of good fits goes down from $42\,\%$ to
$31\,\%$. For the subsample with less than $5\,\GeV$ of
ISR energy it is reduced from $53\,\%$ to
$40\,\%$.
At the same time, the $W^\pm$ mass shift increases by approximately $0.2\,\GeV$
for the whole sample. For the subsample with medium $\EISR$, however,
the mass shift increases from $+1.68\,\GeV$ to $+2.19\,\GeV$.
 
On the other hand, with the $4j+\gamma$ hypothesis, the $5C$ fit performance
is much less affected by beamstrahlung effects:
The fraction of good fits stays almost constant,
and the $\sigma$ of the Gaussian width of the mass peak
increases only moderately, from $1.25\,\GeV$ to $1.35\,\GeV$ for the 
complete sample.
The mass shift increases by approximately $0.2\,\GeV$
for the full sample, which is similar to the $4j$ hypothesis.
However, for the subsample with $5 \GeV \le \EISR \le 30\,\GeV$
the mass shift is significantly reduced from
$+2.19$ to $+1.07\,\GeV$ by the inclusion of the photon in the fit.
The increase of the mass shift with respect to the ISR only case 
indicates that the $4j+\gamma$ hypothesis cannot fully accommodate
beamstrahlung effects,
because typically both beam particles radiate off significant energy.
This may necessitate the inclusion of 
a second photon in the fit.

\begin{figure}[hbt]
   \label{fig:m35x}
   \centering
   \subfigure{
      \epsfig{file=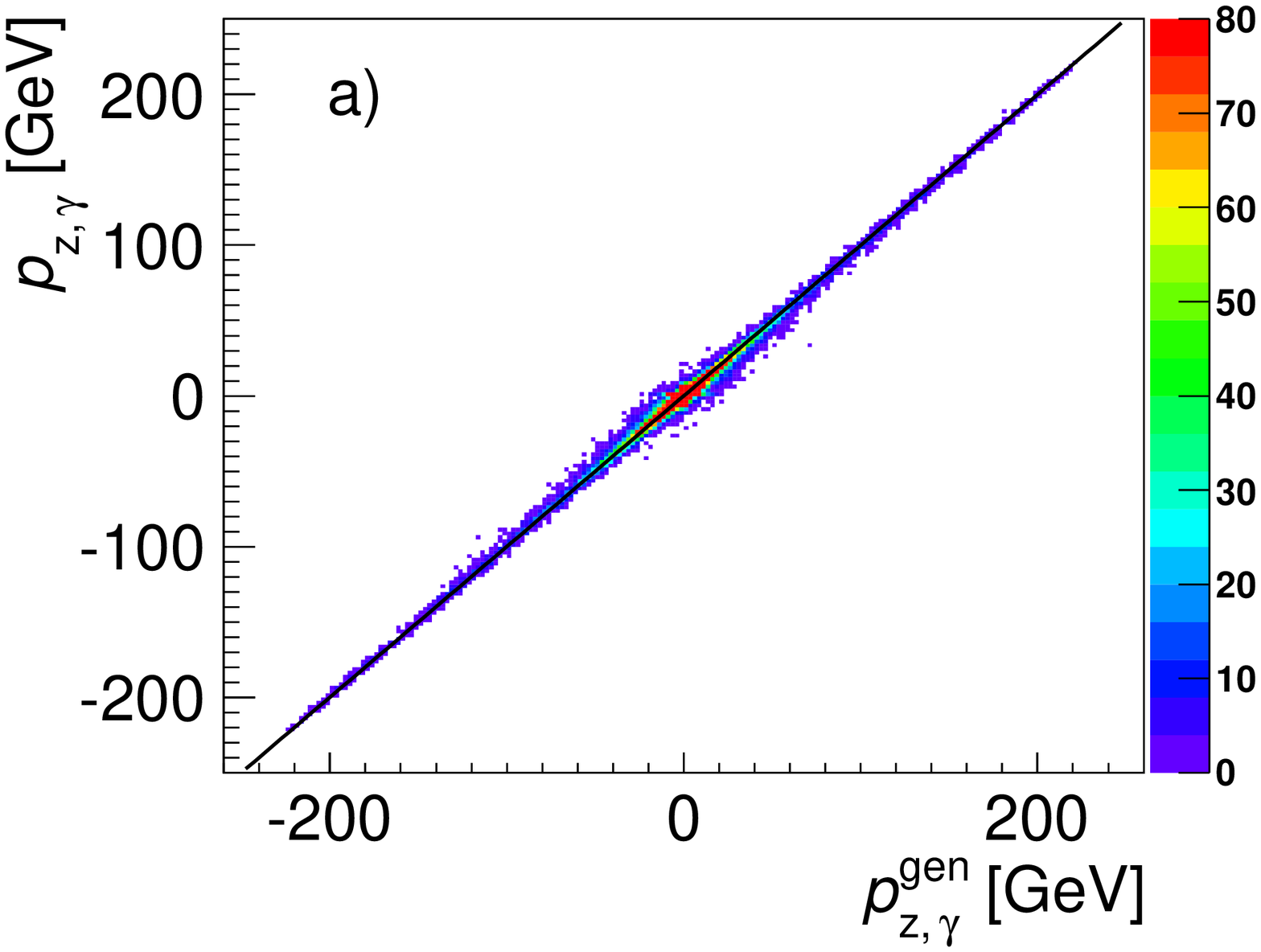,width=0.475\textwidth}
   }
   \subfigure{
      \epsfig{file=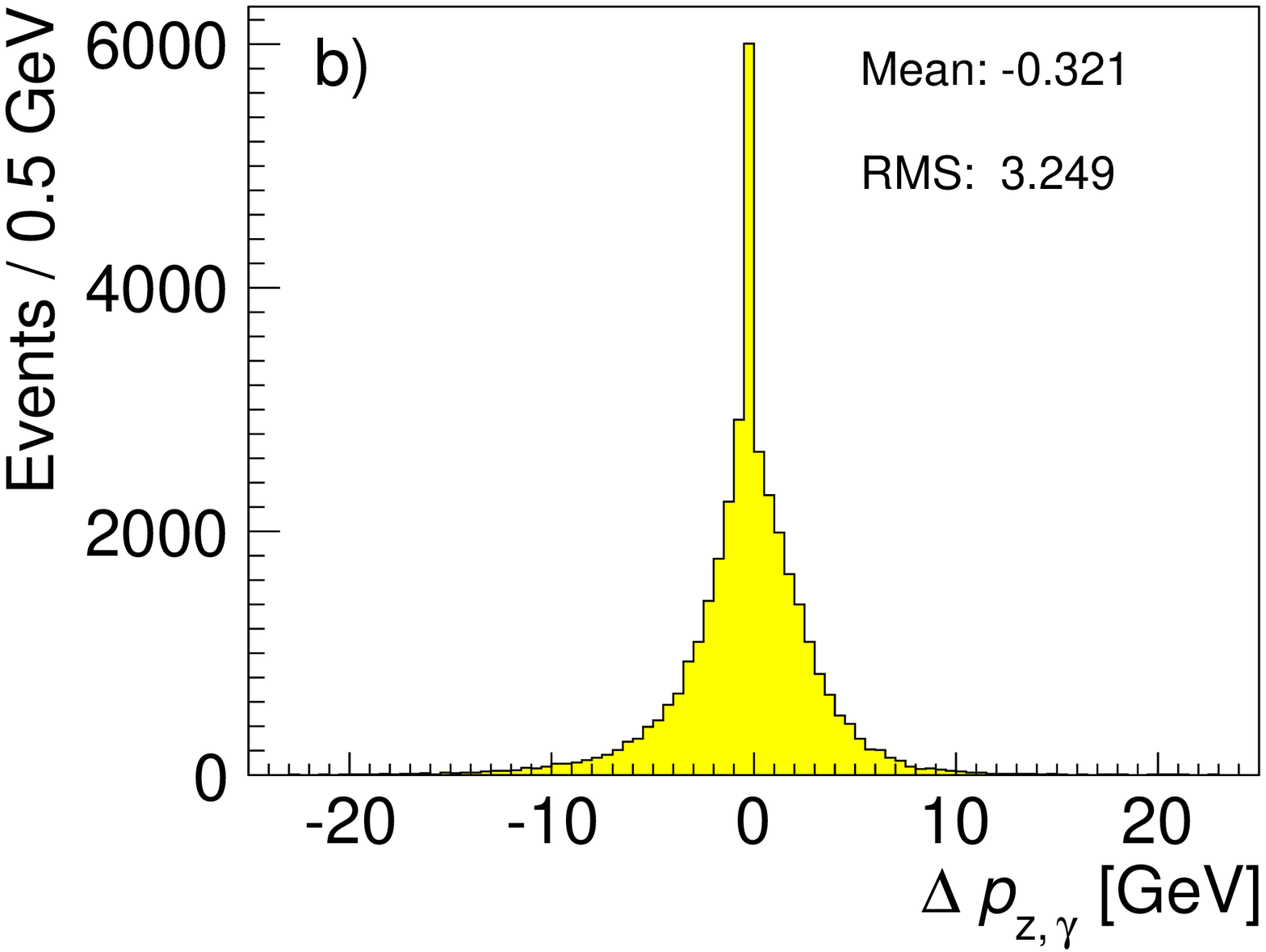,width=0.475\textwidth}
   }
   \caption{Fitted photon momentum $\pzgamma$ plotted against  
        the true momentum $\pzgammagen$ of the most energetic ISR+beamstrahlung
        photon combination in the event (a), and the difference 
        $\Delta \pzgamma = \sign (\pzgamma) \cdot (\pzgamma - \pzgammagen) $ (b).
   \label{fig:pzresoltion}
        }
\end{figure}

As a final check,
Fig.~\ref{fig:pzresoltion} shows the fitted longitudinal momentum $\pzgamma$ 
of the photon versus the generated $\pzgammagen$ of the most energetic ISR+beamstrahlung
photon pair in the event,
where the momenta of the ISR and beamstrahlung photons with either positive or
negative $\pz$ are added.
It can be seen that the fitted photon momentum $\pzgamma$
corresponds quite well to the true momentum, without any visible bias.
In particular, the fact that the photon is treated as having 
a measured $\pzgamma=0$ does not lead to a large bias towards
small values of $\pzgamma$.
This is explained by the fact that the function $\pzgamma\,(\eta)$ of
Eq.~(\ref{eq:parametrization}) rises very rapidly.

The right side of Fig~\ref{fig:pzresoltion} shows the difference
$\Delta \pzgamma = \sign (\pzgamma) \cdot (\pzgamma - \pzgammagen)$.
The mean $\langle \Delta \pzgamma \rangle = -0.32\,\GeV$ is small,
and negative, showing that the reconstructed $|\pzgamma|$ is slightly
smaller on average than the generated one, as expected, but that this
bias is indeed quite small. The resolution for $\pzgamma$ is found to be
$3.25\,\GeV$. 
 
\section{Summary and Conclusions}
\label{sec:summary}

In this paper a method is proposed to take the effect 
of ISR into account in kinematic fits by introducing
a photon  that is treated as if its measured
momentum were zero. 
The longitudinal momentum $\pzgamma$ is expressed as
a function $\pzgamma\,(\eta)$ of the parameter $\eta$
such that the true
value of $\eta$ follows a normal distribution
with zero mean and unit standard deviation.

The performance of this method is evaluated using a sample
of $e^+ e^- \rightarrow u\bar{d}d\bar{u}$ events,
which is dominated by $W^+W^-$ pair production, 
at $\sqrt{s} = 500\,\GeV$.
The sample includes the effects from ISR and beamstrahlung.
It is fully simulated and reconstructed,
using the simulation for the ILD detector at the ILC.
A $5C$ kinematic fit with energy and momentum conservation constraints
and an equal-mass constraint is applied, 
and the results for the fit hypothesis with four jets and a photon
are compared to three alternatives: a $5C$ fit with
a conventional four jet hypothesis,
a $3C$ fit where the energy and longitudinal momentum constraints are 
dropped,
and the results obtained without a kinematic fit.

The $5C$ fit with the new $4j+\gamma$ hypothesis performs
as well as a $5C$ fit with a $4j$ hypothesis in terms of 
resolution, while a $3C$ is significantly worse and
does not yield any improvement over a mass reconstruction
without any kinematic fit.

For events with significant energy
from ISR photons ($5 \GeV \le \EISR \le 30\,\GeV$), 
the fraction of good fits with a fit probability
$p>0.001$ drops from $40\,\%$ to $12\,\%$ for a $5C$ fit
without a photon,
and goes to zero for $\EISR > 30\,\GeV$.
In addition, as the missing energy is distributed to the jets 
by such a fit,
a shift of the reconstructed di-jet masses towards larger values
is observed. 

Both problems are solved by the new $4j+\gamma$ hypothesis: 
even for large
values of $\EISR > 30\,\GeV$, 
the fraction of good fits 
and the di-jet mass resolution are similar to the values obtained
at $\EISR < 5\,\GeV$,
while the mass shift remains small.

In short, under the $4j+\gamma$ hypothesis, 
a $5C$ fit achieves the same resolution
as with a conventional $4j$ fit hypothesis, 
but independent of the amount
of ISR energy, without developing a mass bias,
and with a similar fraction of good fits as a $3C$ fit.

Although the parametrization $\pzgamma\,(\eta)$ was
developed using the momentum spectrum of ISR photons,
the method also performs well in the presence of
beamstrahlung, at least at the moderate level
expected for the nominal parameter set of the ILC.

In a future development the parametrization could be adapted
to include beamstrahlung effects. This may be necessary
in scenarios with enhanced beamstrahlung, such as the ``low power''
parameter set proposed for the ILC, or at CLIC.
We expect that under such conditions 
the addition of a second photon in the fit
would become necessary in order to take into account
the energy loss suffered by both beam particles.

\section*{Acknowledgements}
\addcontentsline{toc}{section}{Acknowledgements}

We would like to thank the ILD simulation production team, in particular F. Gaede, 
S. Aplin, J. Engels and I. Marchesini, for the production of the 
samples of events used in this work,
and T. Barklow for producing the generated
input files. 

We acknowledge the support of the DFG through the SFB (grant SFB 676/1-2006)
and the Emmy-Noether program (grant LI-1560/1-1).


\begin{footnotesize}


\end{footnotesize}


\end{document}